\documentclass[english,pra,manuscript]{revtex4-1}
\usepackage[T1]{fontenc}
\usepackage[utf8]{inputenc}
\usepackage{color}
\usepackage{pmboxdraw}
\usepackage{mathrsfs}
\usepackage{amsmath}
\usepackage{amssymb}
\usepackage{graphicx}
\usepackage{esint}
\usepackage{babel}
\begin{document}

\title{Elliptical orbits in the phase-space quantization}

\author{Leonardo Andreta de Castro\\
Carlos Alexandre Brasil\\
Reginaldo de Jesus Napolitano}

\address{São Carlos Institute of Physics, University of São Paulo~\\
PO Box 369, 13560-970, São Carlos, SP, Brazil.}
\begin{abstract}
The energy levels of hydrogen-like atoms are obtained from the phase-space
quantization, one of the pillars of the old quantum theory, by three
different methods ─ (i) direct integration, (ii) Sommerfeld's original
method, and (iii) complex integration. The difficulties come from
the imposition of elliptical orbits to the electron, resulting in
a variable radial component of the linear momentum. Details of the
calculation, which constitute a recurrent gap in textbooks that deal
with phase-space quantization, are shown in depth in an accessible
fashion for students of introductory quantum mechanics courses.
\end{abstract}
\maketitle

\section{Introduction}

Few expressions are more frustrating to the student than “it is easy
to prove that” followed by a non-trivial result. On the one hand,
when well employed, that is, when the steps necessary to reach the
final result are a natural consequence of what was explained before,
the expression improves the reader's self-confidence by stimulating
the use of creativity to fill the logical and mathematical gaps in
finding the solution. On the other hand, its inadequate use can make
the student lose too much time in a fruitless pursuit of misguided
paths, leaving them eventually discouraged when a simple sentence
or set of references could point them to the right direction.

This article addresses the specific but recurrent inadequate usage
of “it is easy to show that” in many quantum mechanics textbooks\cite{EisbergResnick1,EisbergResnick2,Tomonaga,Bohm,Born,Sommerfeld}
when deriving the energy levels of a hydrogen-like atom.\cite{Sommerfeld}
The problem consists in solving an integral resulting from the phase-space
quantization conditions\cite{Sommerfeld,Wilson,vanderWaerden,Ishiwara,Sommerfeld1,Sommerfeld2}
─ fundamental in the old quantum theory ─ when the electronic orbits
are allowed to be elliptical (as opposed to circular), thus allowing
for non-zero radial component of the momentum.

Specifically, if the electron orbits the nucleus in an elliptical
trajectory, with the radius varying between $r_{\mathrm{max}}$ and
$r_{\mathrm{min}}$, with constant energy $E$ and angular momentum
$L$, the quantization of the radial action will be given by the integral

\begin{equation}
2\int_{r_{\mathrm{min}}}^{r_{\mathrm{max}}}\mathrm{d}r\sqrt{-\frac{L^{2}}{r^{2}}+\frac{2Zme^{2}}{r}+2mE}=n_{r}h,\label{integralBWS}
\end{equation}
where $h$ is Planck's constant, $m$ is the mass of the electron,
$-e$ its charge, and $Z$ the number of protons in the nucleus. $n_{r}$
is a positive integer, the \emph{radial} quantum number. Our aim is
to solve this to find an expression for the energy $E$.

Introductory quantum mechanics textbooks that deal with the old quantum
theory and present Eq. (\ref{integralBWS}) omit the necessary steps
to solve it ─ or provide general guidelines based on properties of
ellipses ─ and then present the final result of the energy. For example,
Eisberg and Resnick's books\cite{EisbergResnick1,EisbergResnick2}
only show the quantization conditions for the angular and linear momenta,
and then simply give a final expression relating the angular momentum
and the axes of the ellipsis to the energy levels. Tomonaga,\cite{Tomonaga}
by his turn, employs the expression ``the integration in Eq. (20.8)
{[}equivalent to our Eq. (\ref{integralBWS}){]} is elementary and
gives” before reproducing the final result. Bohm\cite{Bohm} leaves
the integral as an exercise, and Born\cite{Born} only cites the result
to highlight the remarkable coincidence between this and Dirac's later
results. He later solves the equation in an appendix using the same
method Sommerfeld\cite{Sommerfeld,Sommerfeld1} employed to solve
the integral.

This fact does not constitute a severe hindrance in the learning process,
since Schrödinger's (and Dirac's) equation provides the complete solutions
for the hydrogen atom.\cite{Cohen-Tannoudji,Sakurai} However, the
hydrogen atom is one of the most important elementary systems that
can be solved with quantum mechanics (together with the potential
well and the harmonic oscillator), and we consider relevant that the
students know how to work out the details even in the old quantum
theory. Here, we explore these calculations in detail, so the readers
can focus more of their time on the physical interpretation of the
results.

The article is organized as follows: Sec. 2 provides a brief historical
introduction to the old quantum theory; Sec. 3 shows how the integral
that quantizes the radial component of the action can be computed
without previous knowledge of its orbits; Sec. 4 describes the original
method Arnold Sommerfeld employed to solve the integral using the
properties of an ellipse and a few other tricks. Conclusions and further
historical context are presented in Sec. 5. Appendices present details
of the more difficult integrations.

\section{History Overview}

In the period of time between Max Planck's proposed quantization of
energy in 1900\cite{Studart,Jammer} and the development of the first
forms of matrix\cite{footnoteHeisenberg} and wave\cite{footnoteSchrodinger}
mechanics ─ beginning in 1925 and 1926, respectively ─ the physics
of the atomic phenomena evolved into a theory now known as \emph{old
quantum mechanics }(OQM).\cite{Jammer,Bucher} Analyses of Planck's
reasoning to derive the quantum of action $h$ are beyond the scope
of this article and can be found in the references,\cite{Studart,Jammer,Kuhn,Straumann,Feldens}
along with translations of Planck's original works.\cite{ter Haar,Planck1,Planck2}
Below, we will see how this was a hybrid theory that prescribed a
classical approach to the problem before the quantization could be,
somehow, introduced.

\subsection{Niels Bohr's contributions}

The greatest achievement of the OQM was the explanation of the discrete
emission spectrum of hydrogen, a problem that had been intriguing
physicists as early as the 19th century.\cite{Jammer} The first step
was Niels Bohr's atomic model,\cite{Bohr1,Bohr2,Bohr3,Calouste,Parente,KraghPT,Svidzinsky}
which allowed the theoretical derivation of Balmer's formula,\cite{Banet1,Banet2}
and which was shortly after complemented by Arnold Sommerfeld's work\cite{Sommerfeld1,Sommerfeld2,Sommerfeld3,Sommerfeld4}
that explained the fine structure ─ first observed by Albert Michelson
in 1891.\cite{Jammer}

A common oversight in introductory quantum mechanics textbooks\cite{EisbergResnick1,EisbergResnick2,Tomonaga}
is to introduce Bohr's atomic model by just citing his postulates,
as if they were simple and obvious considerations, when in fact these
were the result of a careful analysis of the spectral data and of
the theoretical toolkit available at the time.\cite{Parente} The
very quantization of the linear momentum is commonly presented as
a postulate, despite it actually being a consequence.

Bohr presented his atomic model in three articles published in the
\emph{Philosophical Magazine }in 1913,\cite{Bohr1,Bohr2,Bohr3,Calouste}
``On the constitution of atoms and molecules.'' In the first one,
he presents two postulates, reproduced literally below:
\begin{enumerate}
\item “That the dynamical equilibrium of the systems in the stationary states
can be discussed by help of the ordinary mechanics, while the passing
of the systems between different stationary states cannot be treated
on that basis.
\item “That the latter process is followed by the emission of a homogeneous
radiation, for which the relation between the frequency and the amount
of energy emitted is the one given by Planck's theory.”
\end{enumerate}
Initially, Bohr, aware that the circular orbits are unstable, makes
a classical energetic analysis of the electron-proton system ─ i.
e., he applies postulate (1). This postulate introduces the \emph{stationary
states }and was responsible for a large share of the criticism towards
his model.\cite{KraghEPJH} Bohr then assumes that Balmer's empirical
formula could be explained in terms of a difference in energies, which
he associates to different orbits. The mechanical frequency of rotation
of the electron around the nucleus and the frequency of emitted radiation
are associated with one another and with Planck's constant ─ here,
postulate (2) is applied. The distinction between frequencies is an
innovative aspect of his model.\cite{KraghEPJH,Darrigol} Bohr's solution
offered the following expression for the energies:

\begin{equation}
E_{n}=-\frac{R_{\mathrm{H}}}{n^{2}},\label{eq:EBohr}
\end{equation}
where $n$ is an integer, the \emph{principal quantum number}, which
specifies a given circular orbit of the electron, and $R_{\mathrm{H}}$
is the \emph{Rydberg constant} of the hydrogen atom.

The greatest achievement of Bohr's model was to derive Balmer's formula
and $R_{\mathrm{H}}$ in terms of fundamental constants:

\[
R_{\mathrm{H}}=\frac{1}{4\pi}\frac{me^{4}}{\hbar^{3}},
\]
where there appeared as well a parameter that would later become of
widespread use, Bohr's radius $r_{1}$ ─ the radius of the orbit that
remains closest to the nucleus:

\[
r_{1}=\frac{\hbar^{2}}{me^{2}}.
\]

In 1922, Niels Bohr was awarded a Nobel prize in Physics ``for his
services in the investigation of the structure of atoms and of the
radiation emanating from them''.\cite{Nobel22}

In his analysis of the hydrogen atom,\cite{Bohr1,Calouste} to find
the aforementioned relation between the rotation and emission frequencies,\cite{Parente}
Bohr employs ─ still before formalizing it ─ one of the forms of the
correspondence principle.\cite{Jammer,Liboff,Whittaker} This principle\cite{Liboff}
(which states that when we take the limits of some basic parameters
of quantum mechanics we must find the classical results again), despite
being vital in OQM, has different formulations due to Planck and Bohr
(and also by Heisenberg, but we will not approach this here), which
are \emph{not }equivalent \cite{Makowski}. In short, according to
Planck, the correspondence principle consisted in the recovery of
the classical result when $\hbar\to0$,\cite{footnoteH} while to
Bohr it corresponds to taking $n\to\infty$, where $n$ is a quantum
number related to the frequencies of the system. Afterwards, Bohr\textcolor{black}{\cite{Liboff}}
would provide a more precise formulation \textcolor{black}{of his
principle:}
\begin{quotation}
\textcolor{black}{``As far as the frequencies are concerned we see
that in the limit where $n$ is large, there exists a close relation
between the ordinary theory {[}classical picture{]} of radiation and
the theory of spectra based on {[}the two quantum postulates{]}.''}
\end{quotation}
In fact, to be universally valid, the principle of correspondence
requires both formulations, with the limits $n\to\infty$ and $h\to0$
linked by the action $J=nh$.\cite{Hassoun} One can treat Bohr's
atom emphasizing Bohr's formulation of the uncertainty principle (related
to $r\to\infty$), instead of using directly the balance of the Coulomb
force and the centripetal force in the electron.\cite{Bhattacharyya}

In 1916, while attempting to quantize the hydrogen atom subject to
external electric fields (Stark effect) or magnetic fields (Zeeman
effect), Sommerfeld noticed that Bohr's rules led to a greater number
of spectral lines than observed experimentally. With the help of the
correspondence principle, Bohr could derive selection rules, which
specified how quantum numbers can vary; and, with Hendrik Kramers's
assistance, he explained the intensity of the lines in the hydrogen
spectrum.\cite{Darrigol}

\textcolor{black}{Correcting the calculation for the reduced mass
of the system with Kramers,\cite{Darrigol} Bohr managed to successfully
explain the spectrum of helium, increasing the acceptance of his theory.\cite{Jammer}
Also remarkable are the spectra of a few molecules,\cite{Bohr3,Calouste}
which agree with experimental data.\cite{KraghPT,Svidzinsky}}

\subsection{Arnold Sommerfeld's formulation}

OQM was based on the following rules:\cite{Jammer}
\begin{itemize}
\item the use of classical mechanics to determine the possible motions of
the system;
\item the imposition of certain quantum conditions to select the actual
or allowed motions;
\item the treatment of radiative processes as transitions between allowed
motions subject to Bohr's frequency formula.
\end{itemize}
During calculation ─ and not just to verify the consistency of the
final solutions ─ an adequate version of the correspondence principle
could be used. 

Analyses of the phase space were already being conducted by Max Planck,\cite{Jammer,Straumann,Eckert}
after all Planck's constant\cite{Studart,Feldens,ter Haar,Planck1,Planck2}
$h$ has dimension of action (and angular momentum). Planck's proposal
was the rule defined by

\[
\int\int\mathrm{d}q\;\mathrm{d}p=h,
\]
where $q$ is a generalized coordinate, and $p$ its corresponding
conjugate momentum.\cite{FetterWalecka} Planck discussed this subject
in the first Solvay Congress,\cite{Jammer,Straumann} while Sommerfeld,
in the same event, presented a postulate according to which ``in
‘every purely molecular process’, the quantity of action

\[
\int_{0}^{\tau}L\:\mathrm{d}t=\hbar
\]
is exchanged, where $\tau$ is the duration of the process and $L$
is the Lagrangian.''\cite{Straumann} This elicited a discussion
with his colleagues, including Einstein. Finally, Sommerfeld presented
to the Bavarian Academy of Sciences the phase-space quantization in
two communications from December, 1915,\cite{Sommerfeld3} and January,
1916.\cite{Sommerfeld4} In the same year he would write a treatise
for the Anallen der Physik\cite{FootnoteSommerfeld} where he detailed
his ideas about the quantization of each degree of freedom for atomic
processes,\cite{Sommerfeld1,Bucher,Sommerfeld2} so that, for $f$
degrees of freedom, we have $f$ equations

\begin{equation}
\oint\mathrm{d}q_{k}\;p_{k}=n_{k}h,\label{eq:BIWS}
\end{equation}
where $q_{k}$ is a generalized coordinate and $p_{k}$ is its corresponding
momentum, $n_{k}$ is a non-negative integer, $k$ varies between
$1$ and $f$, and the integral is extended through a whole period
of $q_{k}$.\cite{Tomonaga,Jammer}

Other three authors have proposed Eq. (\ref{eq:BIWS}): (i) Jun Ishiwara
in 1915;\cite{Ishiwara} (ii) William Wilson in 1916;\cite{Wilson}
(iii) Niels Bohr in 1918.\cite{FootnoteBohr} Here we will focus on
Sommerfeld's contributions, but an overview of these versions can
be found in the references.\cite{vanderWaerden,Jammer}

As mentioned above, Bohr's model treated the atom as having many possible
\emph{circular} electronic orbits around the nucleus, designated by
the principal quantum number $n$, which, for the values $1$, $2$,
$3$ and $4$, corresponded in spectroscopic notation to $s$, $p$,
$d$ and $f$,\cite{FootnoteF} from \textbf{S}ingle, \textbf{P}rincipal,
\textbf{D}iffuse and \textbf{F}undamental ─ this notation was used
by Sommerfeld, among others.\cite{Jammer,Seth1}

Sommerfeld, on the other hand, sought a more abstract interpretation
for the numerical regularities in the spectral lines, a less pictorial
and more formal approach that was even called ``number mysticism.''
\cite{Seth1,Seth2} His colleague Wilhelm Wien in Munich called Sommerfeld's
\emph{Atomistik }an \emph{Atom-Mystik}.\cite{Seth1} This point of
view was synthesized by Pauli:\cite{Seth2,PauliNobel}
\begin{quotation}
``At that time there were two approaches to the difficult problems
connected with the quantum of action. One was an effort to bring abstract
order to the new ideas by looking for a key to translate classical
mechanics and electrodynamics into quantum language which would form
a logical generalization of these. This was the direction which was
taken by Bohr's ‘correspondence principle.’ \emph{Sommerfeld, however,
preferred, in view of the difficulties which blocked the use of the
concepts of kinematical models, a direct interpretation, as independent
of models as possible, of the laws of spectra in terms of integral
numbers, following, as Kepler once did in his investigation of the
planetary system, an inner feeling for harmony. }Both methods, which
did not appear to me irreconcilable, influenced me. The series of
whole numbers $2$, $8$, $18$, $32$... giving the lengths of the
periods in the natural system of chemical elements, was zealously
discussed in Munich, including the remark of the Swedish physicist,
Rydberg, that these numbers are of the simple form $2n^{2}$, if $n$
takes on all integer values. \emph{Sommerfeld tried especially to
connect the number 8 and the number of corners of a cube.}'' {[}italics
ours{]}
\end{quotation}
The highlighted sentences contain the essence of the method employed
by Sommerfeld and his disciples, which he would apply as well to the
Zeeman Effect.\cite{Eckert,Seth2} The analogy with Kepler led Sommerfeld
to consider the general case of \emph{elliptical} orbits (in this
non-literal sense).\cite{Bucher} We will see these calculations in
detail in the next section.

\section{Quantization of the Orbits}

The description of phenomena in the OQM was heavily based on their
classical dynamics. For this reason, the atom would be studied through
the knowledge of classical celestial mechanics, a field that had already
collected many results for a point particle subject to a central force
proportional to the inverse square distance. The Lagrangian formalism
is especially useful for future quantization of the action:

\begin{equation}
\mathcal{L}=K-V=\frac{m\mathbf{\dot{r}}^{2}}{2}+\frac{Ze^{2}}{r},\label{eq:Lagrangian}
\end{equation}
where the electron is at position $\mathbf{r}$, has mass $m$ and
charge $-e$, and is orbiting $Z$ protons at the origin whose mass
is approximated as infinite.\cite{FootnoteMass} $K$ and $V$ represent
kinetic and potential energy.\cite{FootnoteGaussian}

The equation of motion for any given coordinate $q$ in this formulation
is:

\[
\frac{\mathrm{d}}{\mathrm{d}t}\frac{\partial\mathcal{L}}{\partial\dot{q}}=\frac{\partial\mathcal{L}}{\partial q}.
\]
For the Cartesian coordinate $z$ and the Lagrangian (\ref{eq:Lagrangian}),
we have:

\[
m\ddot{z}=-\frac{Ze^{2}}{r^{3}}z.
\]

If we choose $xy$ plane as the one that contains both the initial
position $\mathbf{r}\left(0\right)$ and velocity vector $\dot{\mathbf{r}}\left(0\right)$,
we will have $z\left(0\right)=\dot{z}\left(0\right)=0$, and as a
consequence $\ddot{z}=\dot{z}=z=0$ for every instant of time. For
this reason, the orbit can be described in two dimensions, and we
just need to quantize these two coordinates according to Eq. (\ref{eq:BIWS})
to find the energy levels of the atom:

\begin{eqnarray*}
\oint\mathrm{d}q_{1}\:p_{1} & = & n_{1}h,\\
\oint\mathrm{d}q_{2}\:p_{2} & = & n_{2}h.
\end{eqnarray*}
In accordance to the Lagrangian formalism, $p_{1},p_{2}$ are the
partial derivatives of the Lagrangian with respect to the first time
derivative of these coordinates.

We may choose these two coordinates in any way that seems us fit,
as long as you know how the generalized momenta depend on the generalized
coordinates, so you can proceed to calculate the integrals.

Fortunately, solving the equations of motions may prove unnecessary
for our purposes, which are just of finding the energy levels of the
atom. We may take a shortcut by using the conservation of total energy
$E$ to establish a relationship between the generalized coordinates
and the generalized momenta:

\begin{equation}
E=\frac{m\mathbf{\dot{r}}^{2}}{2}-\frac{Ze^{2}}{r}.\label{eq:TotalEnergy}
\end{equation}
This $E$ must be negative for orbits, otherwise the electron would
not be bound to the nucleus, and free to escape towards infinity.

From Eq. (\ref{eq:TotalEnergy}), one can find $p_{1}$ as a function
of $p_{2},q_{1},q_{2}$. An additional relation may be necessary to
solve the integrals, which may come from a clever manipulation of
the equations of motion:

\begin{eqnarray*}
\dot{p}_{1} & = & \frac{\partial\mathcal{L}}{\partial q_{1}},\\
\dot{p}_{2} & = & \frac{\partial\mathcal{L}}{\partial q_{2}}.
\end{eqnarray*}
The most traditional method of solution would employ the polar coordinates
$r,\theta$. In these coordinates, the Lagrangian (\ref{eq:Lagrangian})
becomes:

\[
\mathcal{L}=\frac{m\dot{r}^{2}}{2}+\frac{mr^{2}\dot{\theta}^{2}}{2}+\frac{Ze^{2}}{r},
\]
yielding the generalized momenta:

\begin{eqnarray*}
p_{r} & = & \frac{\partial\mathcal{L}}{\partial\dot{r}}=m\dot{r},\\
p_{\theta} & = & \frac{\partial\mathcal{L}}{\partial\dot{\theta}}=mr^{2}\dot{\theta}.
\end{eqnarray*}

The angular momentum, henceforth represented by $L$, is a constant,
because the Lagrangian does not depend on $\theta$:

\[
\frac{\mathrm{d}}{\mathrm{d}t}L=\frac{\partial\mathcal{L}}{\partial\theta}=0.
\]
As in a complete orbit the angle $\theta$ goes from $0$ to $2\pi$,
the phase-space quantization condition translates as:

\[
\oint\mathrm{d}\theta\:p_{\theta}=\int_{0}^{2\pi}\mathrm{d}\theta\;L=n_{\theta}h,
\]
or $L=n_{\theta}\hbar$. Fortunately, this also means that the total
energy in Eq. (\ref{eq:TotalEnergy}) can be written just in terms
of $p_{r}$ and $r$:

\begin{equation}
E=\frac{p_{r}^{2}}{2m}+\frac{L^{2}}{2mr^{2}}-\frac{Ze^{2}}{r}.\label{eq:TotalEnergy-1}
\end{equation}
Solving for $p_{r}$, this momentum can be written as a function of
a single variable:

\[
p_{r}\left(r\right)=\pm\sqrt{-\frac{L^{2}}{r^{2}}+\frac{2Zme^{2}}{r}+2mE}.
\]
The sign is chosen according to whether $r$ is increasing (positive)
or decreasing (negative). The points where $p_{r}=0$ are the extremities
of the movement in $r$. They correspond to the roots of the following
second-degree polynomial:

\[
L^{2}\left(\frac{1}{r}\right)^{2}-2Zme^{2}\left(\frac{1}{r}\right)-2mE=0.
\]
Solving it, we find the values of the radius in the periapsis (closest
distance to the nucleus) and apoapsis (farthest distance to the nucleus):

\begin{equation}
\begin{cases}
r_{\mathrm{min}}= & \frac{1}{\frac{Zme^{2}}{L^{2}}+\frac{1}{L}\sqrt{2mE+\frac{Z^{2}m^{2}e^{4}}{L^{2}}}},\\
r_{\mathrm{max}}= & \frac{1}{\frac{Zme^{2}}{L^{2}}-\frac{1}{L}\sqrt{2mE+\frac{Z^{2}m^{2}e^{4}}{L^{2}}}}.
\end{cases}\label{eq:Limits}
\end{equation}

These two extremes become the same when $E=-Z^{2}me^{4}/2\left(n_{\theta}\hbar\right)^{2}$,
which is the case of circular orbits. The energy levels of an elliptical
orbit, however, will also depend on the quantum number of the radius,
$n_{r}$, and its relation with energy can only be found by solving
the integral (equivalent to Eq. \ref{integralBWS}):

\begin{equation}
\oint\mathrm{d}r\:p_{r}=2\int_{r_{\mathrm{min}}}^{r_{\mathrm{max}}}\mathrm{d}r\:\sqrt{-L^{2}\left(\frac{1}{r}-\frac{1}{r_{\mathrm{max}}}\right)\left(\frac{1}{r}-\frac{1}{r_{\mathrm{min}}}\right)}=n_{r}h,\label{eq:IntegralSqrt}
\end{equation}
where we used the fact that the integral through the entire orbit
corresponds to two equal terms from $r_{\mathrm{min}}$ to $r_{\mathrm{max}}$,
and then back from $r_{\mathrm{max}}$ to $r_{\mathrm{min}}$, which
we simplified to a single integral multiplied by a factor $2$. The
sign chosen is positive because in this region $r$ is increasing.

This integral can be calculated in the complex plane with the help
of the residue theorem, which is explained in Appendix A. Here, we
will adopt an approach that requires simpler calculus by simply changing
variables to a $\vartheta$ such that:

\[
r^{-1}=\left(\frac{r_{\mathrm{min}}^{-1}+r_{\mathrm{max}}^{-1}}{2}\right)+\left(\frac{r_{\mathrm{min}}^{-1}-r_{\mathrm{max}}^{-1}}{2}\right)\sin\vartheta.
\]
Clearly, $\vartheta=\pi/2$ when $r=r_{\mathrm{min}}$, and $\vartheta=-\pi/2$
when $r=r_{\mathrm{max}}$. With a little algebra the integral becomes
the following:

\begin{equation}
2L\int_{-\pi/2}^{\pi/2}\mathrm{d}\vartheta\:\frac{\cos^{2}\vartheta}{\left(\varepsilon+\sin\vartheta\right)^{2}}=n_{r}h.\label{eq:Integral}
\end{equation}
where we defined

\begin{equation}
\varepsilon\equiv\frac{r_{\mathrm{min}}^{-1}+r_{\mathrm{max}}^{-1}}{r_{\mathrm{min}}^{-1}-r_{\mathrm{max}}^{-1}}=\frac{1}{\sqrt{1+2mE\left(\frac{L}{Zme^{2}}\right)^{2}}}>1.\label{eq:Epsilon1}
\end{equation}
The integral from Eq. (\ref{eq:Integral}) is found in mathematical
handbooks, and a direct method of solving it that any student of quantum
mechanics can understand can be found in Appendix B. The final result
is

\[
\int_{-\pi/2}^{\pi/2}\mathrm{d}\vartheta\:\frac{\cos^{2}\vartheta}{\left(\varepsilon+\sin\vartheta\right)^{2}}=\pi\frac{\varepsilon}{\sqrt{\left(\varepsilon^{2}-1\right)}}-\pi.
\]

Hence, back to Eq. (\ref{eq:Integral}), we find, at last:

\begin{equation}
2\pi L\left\{ \frac{1}{\sqrt{1-\varepsilon^{-2}}}-1\right\} =n_{r}h.\label{eq:Intermediate}
\end{equation}
From the definition of $\varepsilon$ in Eq. (\ref{eq:Epsilon1}),

\[
\frac{1}{\sqrt{1-\varepsilon^{-2}}}=\frac{1}{\sqrt{-2mE}}\frac{Zme^{2}}{L}.
\]

Knowing this, we can solve Eq. (\ref{eq:Intermediate}) for $E$ to
find the energy levels:

\[
E=-\frac{Z^{2}me^{4}}{2\left(n_{r}\hbar+L\right)^{2}}.
\]
Equivalently, we can replace the angular momentum $L$ by its quantized
values $n_{\theta}\hbar:$\cite{Bohm}

\begin{equation}
E=-\frac{m}{2\hbar^{2}}\frac{Z^{2}e^{4}}{\left(n_{r}+n_{\theta}\right)^{2}}.\label{eq:EnergyLevels}
\end{equation}

These are the energy levels of the elliptical orbits, which are the
same as in the case of circular orbits, except that now the energy
levels are degenerate, because there are two quantum numbers in the
denominator. This also indicates that this result ─ in which we arrived
using only the integrals, and no further information about the form
of the trajectory ─ is correct. Using properties of the ellipse, however,
these procedures can be made a little shorter, as seen in the next
section.

\section{A Few Shortcuts}

In the following two subsections, we will see how to derive the same
answer from Eq. (\ref{eq:EnergyLevels}) using the parametric formula
of the ellipse:

\begin{equation}
r\left(\theta\right)=a\frac{1-\varepsilon^{2}}{1+\varepsilon\cos\theta},\label{eq:ellipse}
\end{equation}
where $a>0$ is the semi-major axis, and $\varepsilon$ is a number
confined to the interval $0<\varepsilon<1$ called eccentricity. Note
that this definition of $\varepsilon$ is different from the one used
in the previous section, but the use of the same Greek letter is useful
due to the similar role both play in the integrand.

Inspecting Eq. (\ref{eq:ellipse}), we can see that its extrema occur
when $r=a\left(1\pm\varepsilon\right)$, which correspond to the $r_{\mathrm{min}}$
and $r_{\mathrm{max}}$ from Eq. (\ref{eq:Limits}):

\[
\begin{cases}
r_{\mathrm{min}}= & a\left(1-\varepsilon\right)=\frac{1}{\frac{Zme^{2}}{L^{2}}+\frac{1}{L}\sqrt{2mE+\frac{Z^{2}m^{2}e^{4}}{L^{2}}}},\\
r_{\mathrm{max}}= & a\left(1+\varepsilon\right)=\frac{1}{\frac{Zme^{2}}{L^{2}}-\frac{1}{L}\sqrt{2mE+\frac{Z^{2}m^{2}e^{4}}{L^{2}}}}.
\end{cases}
\]
Adding or multiplying these two, we find out that:

\begin{equation}
\begin{cases}
r_{\mathrm{min}}r_{\mathrm{max}}= & a^{2}\left(1-\varepsilon^{2}\right)=-\frac{L^{2}}{2mE},\\
\frac{r_{\mathrm{min}}+r_{\mathrm{max}}}{2}= & a=-\frac{Ze^{2}}{2E}.
\end{cases}\label{eq:PropsEllipse}
\end{equation}
This pair of results may prove useful below.

\subsection{Integrating in theta}

The solution of Eq. (\ref{integralBWS}), as performed in the previous
section, despite being proposed as an exercise in certain textbooks,\cite{Bohm}
is not how Sommerfeld quantized the atom in his original 1916 article.\cite{Sommerfeld}
He used the fact that the radial momentum $p_{r}$ will inevitably
be a function of the angle $\theta$ to integrate in the other coordinate
instead:

\[
\oint\mathrm{d}r\:p_{r}=\int_{0}^{2\pi}\mathrm{d}\theta\;\frac{\mathrm{dr}}{\mathrm{d}\theta}p_{r}\left(\theta\right)=n_{r}h.
\]
Invoking the fact that $L=mr^{2}\dot{\theta}$ is a constant of motion,
we can write the whole integrand in terms of $r$:

\[
p_{r}\left(\theta\right)=m\frac{\mathrm{d}r}{\mathrm{d}t}=m\frac{\mathrm{d}r}{\mathrm{d}\theta}\dot{\theta}=\frac{L}{r^{2}}\frac{\mathrm{d}r}{\mathrm{d}\theta}.
\]
Hence, we only need $r$ as a function of $\theta$ to calculate:

\begin{equation}
L\int_{0}^{2\pi}\mathrm{d}\theta\;\left(\frac{1}{r}\frac{\mathrm{d}r}{\mathrm{d}\theta}\right)^{2}=n_{r}h\label{eq:SommerfeldInt}
\end{equation}

This at first may look like an increase in the difficulty of the problem,
because it will require writing $r$ and $p_{r}$ in terms of $\theta$.
But if we know Eq. (\ref{eq:ellipse}), the derivative we need is
just:

\[
\frac{1}{r}\frac{\mathrm{d}r}{\mathrm{d}\theta}=\frac{\varepsilon\sin\theta}{1+\varepsilon\cos\theta}.
\]
Therefore, the integral in Eq. (\ref{eq:SommerfeldInt}) can be simplified
to

\begin{equation}
L\int_{0}^{2\pi}\mathrm{d}\theta\;\frac{\sin^{2}\theta}{\left(\varepsilon^{-1}+\cos\theta\right)^{2}}=n_{r}h.\label{eq:InitInt}
\end{equation}

This trigonometric integral is very similar to Eq. (\ref{eq:Integral})
from the last section, and a practical way of solving it can be found
in Appendix B. In short, the final result is:

\[
\int_{0}^{2\pi}\mathrm{d}\theta\;\frac{\sin^{2}\theta}{\left(\varepsilon^{-1}+\cos\phi\right)^{2}}=2\pi\left\{ \frac{1}{\sqrt{\left(1-\varepsilon^{2}\right)}}-1\right\} ,
\]
or, in Eq. (\ref{eq:InitInt}),

\begin{equation}
1-\varepsilon^{2}=\frac{L^{2}}{\left(n_{r}\hbar+L\right)^{2}}.\label{eq:partialAnswer}
\end{equation}

How do we relate $\left(1-\varepsilon^{2}\right)$ to our known constants?
From Eqs. (\ref{eq:PropsEllipse}), we know that

\[
\left(1-\varepsilon^{2}\right)=-\left(-\frac{2E}{Ze^{2}}\right)^{2}\frac{L^{2}}{2mE}=-\frac{2EL^{2}}{Z^{2}me^{4}}
\]
so that Eq. (\ref{eq:partialAnswer}) yields the final answer we had
found in the previous section by direct integration:

\[
E=-\frac{Z^{2}me^{4}}{2\left(n_{r}\hbar+L\right)^{2}}.
\]
As you can see, both methods are very much analogous. The main difference
is that Sommerfeld's original approach requires a little more geometrical
insights.

\subsection{Integrating in time}

Another shortcut to the final solution employs the virial theorem.\cite{Whittaker}
We begin, once again, from the quantization of the action

\[
\oint\mathrm{d}r\:p_{r}=\oint\mathrm{d}r\:m\dot{r}=n_{r}h,
\]

\[
\oint\mathrm{d}\theta\;L=\oint\mathrm{d}\theta\:mr^{2}\dot{\theta}=n_{\theta}h.
\]

This time, however, instead of solving the integrals separately, we
change variables to time and sum the contributions:

\[
\int_{0}^{T}\mathrm{d}t\;m\left(\dot{r}^{2}+r^{2}\dot{\theta}^{2}\right)=h\left(n_{r}+n_{\theta}\right),
\]
where $T$ is the orbital period.

Now divide both sides by $2T$. From Eq. (\ref{eq:TotalEnergy-1}),
we know that the left-hand side became the time average of the kinetic
energy of the system, which according to the virial theorem, is equal
to the average total energy with sign changed.

Hence, the total energy of the electron in this elliptical orbit,
which is a constant, will be:

\begin{equation}
E=-\frac{1}{2T}\int_{0}^{T}\mathrm{d}t\;m\left(\dot{r}^{2}+r^{2}\dot{\theta}^{2}\right)=-\frac{1}{2T}h\left(n_{r}+n_{\theta}\right).\label{eq:E_partial}
\end{equation}

Now, we just have to establish a relationship between the period $T$
and other known parameters. To do this, we invoke conservation of
linear momentum (or Kepler's second law), which states that $L\mathrm{d}t=mr^{2}\mathrm{d}\theta$.
Integrating over the whole orbit, the area integral becomes the total
area of the ellipse, $\pi a^{2}\sqrt{1-\varepsilon^{2}}$:

\[
\frac{1}{2}LT=m\pi a^{2}\sqrt{1-\varepsilon^{2}}.
\]
From Eqs. (\ref{eq:PropsEllipse}), we can rewrite $a$ and $1-\varepsilon^{2}$
in terms of other constants:

\[
a^{2}\sqrt{1-\varepsilon^{2}}=\left(-\frac{Ze^{2}}{2E}\right)\sqrt{-\frac{L^{2}}{2mE}}=-\frac{Ze^{2}}{2E}\frac{L}{\sqrt{-2mE}},
\]
thus yielding

\[
\frac{1}{2T}=\frac{E}{2\pi}\frac{\sqrt{-2mE}}{Zme^{2}}.
\]

Returning to Eq. (\ref{eq:E_partial}), we find

\[
E=-\frac{Z^{2}me^{4}}{2\hbar^{2}\left(n_{r}+n_{\theta}\right)^{2}},
\]
which are once again the energy levels we were looking for.

\section{Further Developments}

As seen above, Sommerfeld quantized separately the linear and angular
momenta, associating them with the quantum numbers $n_{r}$ and $n_{\theta}$.
This results again in Bohr's formula, Eq. (\ref{eq:EBohr}), but this
time

\[
n=n_{r}+n_{\theta}.
\]
The value $n=0$ is excluded \emph{ad hoc} for being unphysical.\cite{Bucher}

In Sec. 3 we observed that different combinations of $n_{r}$ and
$n_{\theta}$ can result in the same $n$, which defines the energy
of the orbit. Therefore, Sommerfeld's model introduces \emph{degenerate
states }for the energy and reaches Bohr's 1913 result through an alternative
path. However, there still lingered the problem of fine structure,
which required one further step.

Attempting to find an expression that included these lines as well,
Sommerfeld proposed a three-dimensional quantization\cite{Jammer,Bucher}
that eliminated the confinement of orbits to a single plane. Moreover,\cite{Sommerfeld2}
using the relativistic formula for the mass at a given velocity $v$:

\[
m=\frac{m_{0}}{\sqrt{1-\left(v/c\right)^{2}}},
\]
where $m_{0}$ is the rest mass of the electron, Sommerfeld would
find a formula that explained the fine structure of the hydrogen spectrum,\cite{Jammer,Bucher}
writing the fine structure constant also in terms of three fundamental
constants:

\[
\alpha=\frac{e^{2}}{\hbar c}
\]
and becoming the first to see that this constant could be written
as the ratio between the orbital velocity of the electron in the closest
orbit and the speed of light.\cite{Kragh1,Kragh2,Kragh3}

A simple relativistic treatment of Bohr's atom (with circular orbits)
results in an energy correction involving terms with $\alpha$, but
these values do not differ significantly from the original non-relativistic
treatment.\cite{Terzis} Thus Sommerfeld's theory acquired real importance,
becoming the highest achievement of OQM,\cite{Bucher} and a confirmation
of special relativity. In his Nobel lecture,\cite{PlanckLecture}
Planck would speak of it with clear enthusiasm:
\begin{quotation}
``...A. Sommerfeld showed that from a logical extension of the laws
of quantum distribution in systems with several degrees of freedom,
and out of consideration of the variability of the inertial mass in
accordance with the relativity theory, that magic formula {[}Balmer's
series formula{]} arose before which both the hydrogen and the helium
spectrum had to reveal the riddle of their fine structure, to such
an extent that the finest present-day measurements, those of F. Paschen,
could be explained generally through it ─\emph{ an achievement fully
comparable with that of the famous discovery of the planet Neptune
whose existence and orbit was calculated by Leverrier before the human
eye had seen it.}'' {[}italics ours{]}
\end{quotation}
In 1929, Arthur Eddington noticed that the value of $\alpha$ could
be written as approximately the inverse of an integer: first $\alpha^{-1}\sim136$,
then $\alpha^{-1}\sim137$. This numerical aspect and the fact that
$\alpha$ can be written as a dimensionless combination of fundamental
constants led to an $\alpha$-numerology, a search for deeper meanings
for $\alpha$.\cite{Kragh1,Kragh2} Curiously, in 1950, the annihilation
of a positron by an electron and the resulting emission of two photons
was observed with a precision of $137^{-1}$.\cite{Benedetti,Machado}
The relativistic treatment for circular orbits sets an upper limit
for the atomic number at $Z<\alpha^{-1}\sim137$,\cite{Terzis} which
as of today has not been violated. What would Dr. Eddington have to
say about that?

Another success of the OQM, accomplished by Sommerfeld and his assistant
Hendrik Kramers, was the initial explanation of the Stark effect.\cite{Jammer,Duncan1,Duncan2}
In this case, despite the very good agreement with the experimental
data available at the time, even Johannes Stark himself showed reservations
towards the theory.\cite{Duncan2,Kragh3} An important role was fulfilled
by Karl Schwarzchild, who made the link between the quantization of
the phase integral (\ref{eq:BIWS}) and the action-variables of the
Hamilton-Jacobi theory.\cite{Jammer,Eckert,FetterWalecka,Duncan1,Duncan2}
This was a natural extension for Schwarzchild, as the action-angle
variables and the Hamilton-Jacobi theory are tools much used in celestial
mechanics, a field he was most familiar with.\cite{FootnoteSchwarzchild}

We see, then, that the OQM brought advances in many problems, but
its gaps would soon be brought to the fore. The theory did not take
into account:\cite{Bucher} (i) the \emph{spin}, necessary for the
complete treatment of the total angular momentum of the electron;
(ii) the \emph{wave nature of particles}, necessary to take into account
its probabilistic character;\cite{MacKinnon,deBroglieSelected,Ludwig}
and (iii) the indistinguishability of particles under certain circumstances.
Points (i) and (ii) are directly related to the \emph{Pauli exclusion
principle}.\cite{EisbergResnick1,Cohen-Tannoudji,Jammer} In the case
of the Stark effect,\cite{Duncan1,Duncan2} for example, the explanation
involved \emph{ad hoc }rules to select the electronic orbits that
best fit the experimental data, and the results depended on the system
of coordinates on which these conditions were imposed.

How could the result of the fine structure of the spectrum given by
the OQM, a heuristic theory with its many gaps, be the same as the
one provided by the modern quantum mechanics, after all the contributions
of Heisenberg, Schrödinger and Dirac? This is due to the symmetry
involved in Sommerfeld's hypothesis of the elliptical orbits and the
use of the relativistic mass, which correspond, in modern quantum
mechanics, to considering relativistic effects and adding the spin-orbit
interaction.\cite{Biedenharn,Vickers} 

Arnold Sommerfeld had great influence also in the consolidation of
modern quantum mechanics, and among his students\cite{Jammer} we
can mention Isidor Isaac Rabi, Edward Uhler Condon, Peter Debye, Wolfgang
Pauli, Werner Heisenberg, Hans Bethe and Alfred Landé,\cite{Jammer,Seth1,Seth2,SomAJP}
the first four laureates of the Nobel prize in Physics. Ironically,
Sommerfeld did not receive such an honor.

His curious teaching style was summarized by himself as:
\begin{quotation}
``Personal instruction in the highest sense of the word is best based
on intimate personal acquaintanceship. Ski trips with my students
offered the best opportunity for that.''
\end{quotation}
His book \emph{Atombau und Spektrallinien\cite{Sommerfeld} }would
be known as ``the Bible of quantum spectroscopy'' and is said to
have impressed David Hilbert, Max Born, and Pieter Zeeman.\cite{Seth1,Duncan1}

His more formal and abstract approach would end up influencing Bohr
himself, who would develop a merely symbolic interpretation of quantum
mechanics in which the classical orbits did not represent the actual
movement of the electrons. But his ideas went further beyond, and
the quantization condition Eq. (\ref{eq:BIWS})

\[
\oint\mathrm{d}q\;p=nh
\]
would influence Heisenberg\cite{Darrigol} in deriving the commutation
relation for the operators $\hat{q}$, $\hat{p}$:

\[
\left[\hat{q},\hat{p}\right]=i\hbar.
\]

Despite being an incomplete theory of strong heuristic character,
the OQM allowed for the first application to the atom of the Planck's
hypothesis concerning the quantum of light, and allowed an intuitive
understanding of the problem. Due to this historical importance, we
hope that in clarifying one of the aspects of the OQM we have helped
not only researchers interested in the early development of quantum
mechanics, but also instructors who can use this paper as an instructional
aid to assist their students in navigating the possible solutions
to this problem.

\section*{Acknowledgements }

L. A. C. acknowledges support from the Coordenação de Aperfeiçoamento
de Pessoal de Ensino Superior (CAPES). C. A. B. thanks Karla Pelogia
(Fakultät 9/Universität Stuttgart), Thaís Victa Trevisan (IFGW/UNICAMP)
and Amit Hagar (HPSM/Indiana University Bloomington) for their help
in locating some of the references used in this paper. R. d. J. N.
thanks Lia M. B. Napolitano for the valuable discussions during the
calculations.

\appendix

\section{Complex Integration}

Here we will consider how to integrate in the complex plane an integral
of the kind found in Eq. (\ref{eq:IntegralSqrt}),\cite{ter Haar,Born-1}
that is, 
\begin{eqnarray}
\mathscr{I} & = & 2\int_{r_{\mathrm{min}}}^{r_{\mathrm{max}}}\mathrm{d}r\,\sqrt{\left(\frac{1}{r_{\mathrm{min}}}-\frac{1}{r}\right)\left(\frac{1}{r}-\frac{1}{r_{\mathrm{max}}}\right)},\label{eq:Reg01}
\end{eqnarray}
which can also be split in two terms as follows: 
\begin{eqnarray}
\mathscr{I} & = & \int_{r_{\mathrm{min}}}^{r_{\mathrm{max}}}\mathrm{d}r\,\sqrt{\left(\frac{1}{r_{\mathrm{min}}}-\frac{1}{r}\right)\left(\frac{1}{r}-\frac{1}{r_{\mathrm{max}}}\right)}+\int_{r_{\mathrm{max}}}^{r_{\mathrm{min}}}\mathrm{d}r\,\left[-\sqrt{\left(\frac{1}{r_{\mathrm{min}}}-\frac{1}{r}\right)\left(\frac{1}{r}-\frac{1}{r_{\mathrm{max}}}\right)}\right].\label{eq:Reg02}
\end{eqnarray}

In the complex plane, the variable of integration is $z=\left|z\right|e^{i\theta}$,
with the phase $\theta$ in the interval $\left[0,2\pi\right)$. Then,
we can think of the two terms on the right-hand side of Eq. (\ref{eq:Reg02})
as resulting from the complex integral:
\begin{eqnarray}
\mathscr{I} & = & \oint_{C}\mathrm{d}z\,f\left(z\right),\label{eq:Reg04}
\end{eqnarray}
where $C$ is the contour of Fig. 1 in the limit in which $\varepsilon\rightarrow0^{+}$
and $\delta\rightarrow0^{+},$ with the function $f\left(z\right)$
defined in such a way that, in this limit, it tends to
\[
-\sqrt{\left(\frac{1}{r_{\mathrm{min}}}-\frac{1}{r}\right)\left(\frac{1}{r}-\frac{1}{r_{\mathrm{max}}}\right)}
\]
from the first and second quadrants of the complex plane, and it tends
to
\[
\sqrt{\left(\frac{1}{r_{\mathrm{min}}}-\frac{1}{r}\right)\left(\frac{1}{r}-\frac{1}{r_{\mathrm{max}}}\right)}
\]
from the third and fourth quadrants.

\begin{figure}
\includegraphics[scale=0.2]{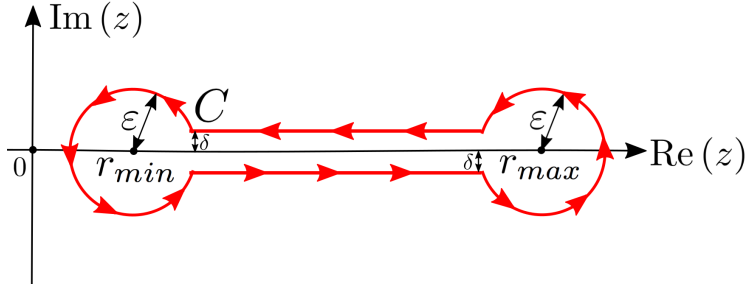}

\caption{Path of the integration of Eq. (\ref{eq:Reg04}) in the complex plane.}
\end{figure}

We can see that $f\left(z\right)$ is discontinuous when $z$ crosses
the segment $\left[r_{\mathrm{min}},r_{\mathrm{max}}\right]$, which
we will call a ``cut line.'' Nevertheless, $f\left(z\right)$ can
be proven\cite{Arfken,Butkov} to be continuous everywhere else in
the complex plane, except at the origin, where it is singular. The
domain of definition of $f\left(z\right)$ in the complex plane comprises
all the complex numbers except the cut line and zero. In its domain,
$f\left(z\right)$ is analytical.

\begin{figure}
\includegraphics[scale=0.2]{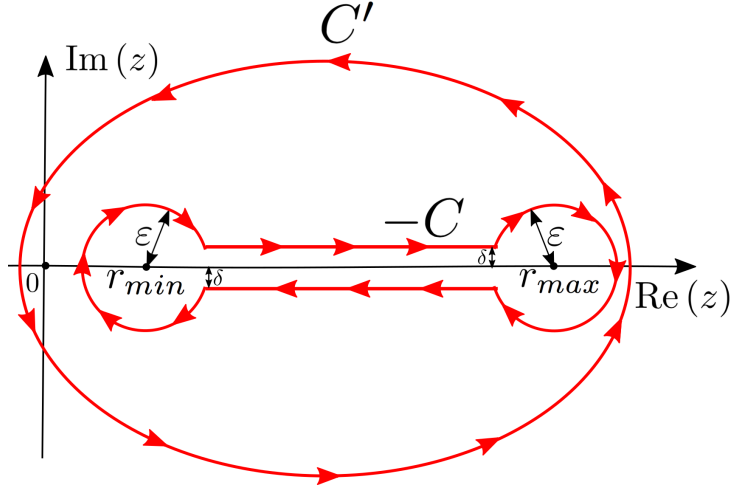}

\caption{Paths $-C$ and $C^{\prime}$ present in Eq. (\ref{eq:Reg05}). Here,
the contour represented as $-C$ is simply the same contour $C$ from
Fig. 1, but traversed in the clockwise direction.}
\end{figure}

Now, consider the contours of Fig. 2. According to the residue theorem,
the integral of $f\left(z\right)$ along the contour $C^{\prime}\cup\left(-C\right)$
is given by
\begin{eqnarray}
\oint_{C^{\prime}\cup\left(-C\right)}\mathrm{d}z\,f\left(z\right) & = & 2\pi i\mathrm{Res}_{z=0}\left[f\left(z\right)\right],\label{eq:Reg05}
\end{eqnarray}
where $\mathrm{Res}_{z=0}\left[f\left(z\right)\right]$ denotes the
residue of $f\left(z\right)$ at zero. It is easy to see that the
integral of Eq. (\ref{eq:Reg05}) can be expressed as the sum of two
other integrals:

\begin{equation}
\oint_{C^{\prime}\cup\left(-C\right)}\mathrm{d}z\,f\left(z\right)=\oint_{C^{\prime}}\mathrm{d}z\,f\left(z\right)+\oint_{-C}\mathrm{d}z\,f\left(z\right)=\oint_{C^{\prime}}\mathrm{d}z\,f\left(z\right)-\oint_{C}\mathrm{d}z\,f\left(z\right).\label{eq:Reg06}
\end{equation}
In the limit in which $\varepsilon\rightarrow0^{+}$ and $\delta\rightarrow0^{+},$
Eqs. (\ref{eq:Reg04}), (\ref{eq:Reg05}) and (\ref{eq:Reg06}) yield
\begin{eqnarray}
\mathscr{I} & = & \oint_{C^{\prime}}\mathrm{d}z\,f\left(z\right)-2\pi i\mathrm{Res}_{z=0}\left[f\left(z\right)\right].\label{eq:Reg07}
\end{eqnarray}
Thus, if we can calculate the integral of $f\left(z\right)$ on the
contour $C^{\prime},$ we can find$\mathscr{I}.$

Since $f\left(z\right)$ is analytical everywhere outside the region
bounded by the contour $C^{\prime},$ it can be expanded in a Laurent
series about the origin $f\left(z\right)=\sum_{n=-\infty}^{+\infty}a_{n}z^{n}$
where the coefficients are given by:
\begin{eqnarray}
a_{n} & = & \frac{1}{2\pi i}\oint_{C^{\prime}}\mathrm{d}z\,\frac{f\left(z\right)}{z^{n+1}},\label{eq:Reg09}
\end{eqnarray}
for all $n\in\mathbb{Z}.$ Integrating both sides of the Laurent series
on $C^{\prime}$ gives:
\begin{eqnarray}
\oint_{C^{\prime}}\mathrm{d}z\,f\left(z\right) & = & \oint_{C^{\prime}}\mathrm{d}z\,\sum_{n=-\infty}^{+\infty}a_{n}z^{n}=a_{-1},\label{eq:Reg10}
\end{eqnarray}
since $\oint_{C^{\prime}}\mathrm{d}z\,z^{n}$ vanishes for all $n\in\mathbb{Z},$
except for $n=-1.$ We can also obtain the same result from the series
$\sum_{n=-\infty}^{+\infty}a_{n-2}w^{-n}$, namely,
\begin{eqnarray}
\oint_{C_{0}}\mathrm{d}w\,\sum_{n=-\infty}^{+\infty}a_{n-2}w^{-n} & = & a_{-1},\label{eq:Reg13}
\end{eqnarray}
where $w\in\mathbb{C}$ and $C_{0}$ is a circular contour centered
at the origin of the complex plane, in the limit when its radius tends
to zero. From Eqs. (\ref{eq:Reg10}) and (\ref{eq:Reg13}), we obtain:

\begin{equation}
\oint_{C^{\prime}}\mathrm{d}z\,f\left(z\right)=\oint_{C_{0}}\mathrm{d}w\,\sum_{n=-\infty}^{+\infty}a_{n-2}w^{-n}=\oint_{C_{0}}\mathrm{d}w\,\frac{1}{w^{2}}\sum_{n=-\infty}^{+\infty}a_{n}w^{-n}.\label{eq:Reg15}
\end{equation}
Noticing the sum on the right-hand side of Eq. (\ref{eq:Reg15}) is
a Laurent series expansion of $f\left(w^{-1}\right)$, we can rewrite
it as 
\begin{eqnarray}
\oint_{C^{\prime}}\mathrm{d}z\,f\left(z\right) & = & \oint_{C_{0}}\mathrm{d}w\,\frac{1}{w^{2}}f\left(\frac{1}{w}\right).\label{eq:Reg16}
\end{eqnarray}
It is clear that the function $f\left(1/w\right)/w^{2}$ can only
have a singularity at the origin, if any, within the region bounded
by the closed curve $C_{0}.$ Thus, according to the residue theorem,
Eq. (\ref{eq:Reg16}) gives:
\begin{eqnarray}
\oint_{C^{\prime}}\mathrm{d}z\,f\left(z\right) & = & 2\pi i\mathrm{Res}_{w=0}\left[\frac{1}{w^{2}}f\left(\frac{1}{w}\right)\right].\label{eq:Reg17}
\end{eqnarray}
Usually, one defines the extended complex plane as the usual plane
and the point at infinity,\cite{Butkov} such that the residue of
Eq. (\ref{eq:Reg17}) can be rewritten as the residue at infinity
of the function $f\left(z\right):$
\begin{eqnarray}
\mathrm{Res}_{z=\infty}\left[f\left(z\right)\right] & \equiv & \mathrm{Res}_{w=0}\left[-\frac{1}{w^{2}}f\left(\frac{1}{w}\right)\right].\label{eq:Reg18}
\end{eqnarray}
Hence, substituting Eqs. (\ref{eq:Reg17}) and (\ref{eq:Reg18}) into
Eq. (\ref{eq:Reg07}), we obtain:
\begin{eqnarray}
\mathscr{I} & = & -2\pi i\mathrm{Res}_{z=\infty}\left[f\left(z\right)\right]-2\pi i\mathrm{Res}_{z=0}\left[f\left(z\right)\right].\label{eq:Reg19}
\end{eqnarray}
With the concept of an extended complex plane, we can think of Eq.
(\ref{eq:Reg19}) as simply the integration of $f\left(z\right)$
on the contour $-C,$ as defined in Fig. 2. It is as if the function
$f\left(z\right)$ had only two singularities outside of the region
bounded by $-C:$ at zero and at infinity. Let us now proceed with
the calculation of both residues.

First, let us calculate the residue at zero. For $r\in\left(r_{\mathrm{min}},r_{\mathrm{max}}\right),$
let us write:
\[
\sqrt{\left(\frac{1}{r_{\mathrm{min}}}-\frac{1}{r}\right)\left(\frac{1}{r}-\frac{1}{r_{\mathrm{max}}}\right)}=\frac{1}{\sqrt{r_{\mathrm{min}}r_{\mathrm{max}}}}\frac{\sqrt{\left(r-r_{\mathrm{min}}\right)\left(r_{\mathrm{max}}-r\right)}}{r}.
\]
Since $f\left(z\right)$ is continuous in the neighborhood of the
origin, we consider $z$ approaching zero from the first quadrant.
With a little algebra it is possible to write:

\begin{equation}
f\left(z\right)=i\exp\left(i\frac{\phi_{\mathrm{min}}+\phi_{\mathrm{max}}}{2}\right)\frac{1}{\sqrt{r_{\mathrm{min}}r_{\mathrm{max}}}}\frac{\sqrt{\left|z-r_{\mathrm{min}}\right|\left|z-r_{\mathrm{max}}\right|}}{z},\mathrm{Re}\left(z\right),\mathrm{Im}\left(z\right)>0\label{eq:Reg20}
\end{equation}
where $\phi_{\mathrm{min}}$ and $\phi_{\mathrm{max}}$ are defined
as the phases of $z-r_{\mathrm{min}}$ and $z-r_{\mathrm{max}}$,
respectively:

\begin{equation}
\begin{cases}
\left(z-r_{\mathrm{min}}\right) & =\left|z-r_{\mathrm{min}}\right|\exp\left(i\phi_{\mathrm{min}}\right),\\
\left(z-r_{\mathrm{max}}\right) & =\left|z-r_{\mathrm{max}}\right|\exp\left(i\phi_{\mathrm{max}}\right).
\end{cases}\label{eq:Reg2122}
\end{equation}
From Eqs. (\ref{eq:Reg2122}), we easily see that, in the limit in
which $z$ approaches, from the first quadrant, a point $r$ in the
real axis within the interval $\left(r_{\mathrm{min}},r_{\mathrm{max}}\right),$
then $\theta\rightarrow0^{+},$ $\phi_{\mathrm{min}}\rightarrow0,$
and $\phi_{\mathrm{max}}\rightarrow\pi.$ This justifies our choice
of sign in Eq. (\ref{eq:Reg20}), because in the first quadrant we
defined $f\left(z\right)$ as having negative sign: 
\begin{eqnarray*}
f\left(z\right) & \rightarrow & \frac{i\exp\left(i\frac{\pi}{2}\right)}{\sqrt{r_{\mathrm{min}}r_{\mathrm{max}}}}\frac{\sqrt{\left|r-r_{\mathrm{min}}\right|\left|r-r_{\mathrm{max}}\right|}}{r}=-\sqrt{\left(\frac{1}{r_{\mathrm{min}}}-\frac{1}{r}\right)\left(\frac{1}{r}-\frac{1}{r_{\mathrm{max}}}\right)}.
\end{eqnarray*}

From Eq. (\ref{eq:Reg20}), it follows that

\begin{eqnarray}
\mathrm{Res}_{z=0}\left[f\left(z\right)\right] & = & \lim_{z\rightarrow0}\left[zf\left(z\right)\right]=i\exp\left(i\pi\right)=-i,\label{eq:Reg27}
\end{eqnarray}
since, using Eqs. (\ref{eq:Reg2122}), $\phi_{\mathrm{min}}\rightarrow\pi$
and $\phi_{\mathrm{max}}\rightarrow\pi$ as $z\rightarrow0$ from
the first quadrant.

Let us use Eq. (\ref{eq:Reg20}) to calculate the residue at infinity
by making $\left|z\right|\equiv R\in\mathbb{R}$ approach infinity
with $\theta\rightarrow0^{+},$ that is, keeping $z$ on the real
axis. From Eqs. (\ref{eq:Reg2122}) we see that, in this limit, $\phi_{min}+\phi_{max}\rightarrow0.$
Hence,
\begin{eqnarray}
z^{2}f\left(z\right) & = & \frac{iR}{\sqrt{r_{\mathrm{min}}r_{\mathrm{max}}}}\sqrt{\left(R-r_{\mathrm{min}}\right)\left(R-r_{\mathrm{max}}\right)}.\label{eq:Reg28}
\end{eqnarray}
Let $w=1/z=1/R.$ If $w\rightarrow0,$ then $R\rightarrow\infty.$
Eq. (\ref{eq:Reg28}) then becomes:

\begin{eqnarray*}
\frac{1}{w^{2}}f\left(\frac{1}{w}\right) & = & \frac{i}{w^{2}\sqrt{r_{\mathrm{min}}r_{\mathrm{max}}}}\sqrt{1-\left(r_{\mathrm{min}}+r_{\mathrm{max}}\right)w+w^{2}},
\end{eqnarray*}
that is,
\begin{eqnarray}
\frac{1}{w^{2}}f\left(\frac{1}{w}\right) & = & \frac{i}{\sqrt{r_{\mathrm{min}}r_{\mathrm{max}}}}\left[\frac{1}{w^{2}}-\frac{1}{2}\left(r_{\mathrm{min}}+r_{\mathrm{max}}\right)\frac{1}{w}+\frac{1}{2}+\dots\right].\label{eq:Reg29}
\end{eqnarray}
The residue is the coefficient of $1/w,$ that is, from Eqs. (\ref{eq:Reg18})
and (\ref{eq:Reg29}):
\begin{eqnarray}
\mathrm{Res}_{z=\infty}\left[f\left(z\right)\right] & = & \mathrm{Res}_{w=0}\left[-\frac{1}{w^{2}}f\left(\frac{1}{w}\right)\right]=i\frac{\left(r_{\mathrm{min}}+r_{\mathrm{max}}\right)}{2\sqrt{r_{\mathrm{min}}r_{\mathrm{max}}}}.\label{eq:Reg30}
\end{eqnarray}
Hence, substituting Eqs. (\ref{eq:Reg27}) and (\ref{eq:Reg30}) into
Eq. (\ref{eq:Reg19}), we obtain:
\begin{eqnarray*}
\mathscr{I} & = & 2\pi\left[\frac{\left(r_{\mathrm{min}}+r_{\mathrm{max}}\right)}{2\sqrt{r_{\mathrm{min}}r_{\mathrm{max}}}}-1\right].
\end{eqnarray*}

According to the values of the sum and product of $r_{\mathrm{min}}$
and $r_{\mathrm{max}}$ given in Eq. (\ref{eq:PropsEllipse}), and
the definition of $\mathscr{I}$ given in Eq. (\ref{eq:Reg01}), this
is equivalent to stating:

\begin{equation}
2\int_{r_{\mathrm{min}}}^{r_{\mathrm{max}}}\mathrm{d}r\,\sqrt{L^{2}\left(\frac{1}{r_{\mathrm{min}}}-\frac{1}{r}\right)\left(\frac{1}{r}-\frac{1}{r_{\mathrm{max}}}\right)}=L\mathscr{I}=2\pi L\left[-\frac{Ze^{2}}{2E}\frac{\sqrt{-2mE}}{L}-1\right].\label{eq:RegXX}
\end{equation}
Equating Eq. (\ref{eq:RegXX}) to $n_{r}h$, as in Eq. (\ref{eq:IntegralSqrt}),
we find the same energy levels of Eq. (\ref{eq:EnergyLevels}):

\[
E=-\frac{Z^{2}me^{4}}{2\left(n_{r}\hbar+L\right)^{2}}.
\]

\section{Solving the trigonometric integrals}

In Sec. 3, we have to solve an integral of the form

\begin{equation}
\int_{-\pi/2}^{\pi/2}\mathrm{d}\vartheta\:\frac{\cos^{2}\vartheta}{\left(\varepsilon+\sin\vartheta\right)^{2}}.\label{eq:IntSec3}
\end{equation}
In Sec. 4, we have another integral that looks almost the same as
the previous one:

\begin{equation}
\int_{0}^{2\pi}\mathrm{d}\theta\;\frac{\sin^{2}\theta}{\left(\varepsilon^{-1}+\cos\theta\right)^{2}}.\label{eq:IntSec4}
\end{equation}
Actually, if we change variables in (\ref{eq:IntSec3}) to $\phi=\pi/2-\vartheta$,
we notice that both are remarkably similar:

\[
\int_{-\pi/2}^{\pi/2}\mathrm{d}\vartheta\:\frac{\cos^{2}\vartheta}{\left(\varepsilon+\sin\vartheta\right)^{2}}=\int_{0}^{\pi}\mathrm{d}\theta\:\frac{\sin^{2}\theta}{\left(\varepsilon+\cos\theta\right)^{2}}.
\]

We may save some time, therefore, if we solve a more general integral
that encapsulates the results of both (\ref{eq:IntSec3}) and (\ref{eq:IntSec4}):

\begin{equation}
\int_{0}^{N\pi}\mathrm{d}\theta\;\frac{\sin^{2}\theta}{\left(\varepsilon+\cos\theta\right)^{2}},\label{eq:IntGeral}
\end{equation}
where $N>0$ is a positive integer. This is still an integral that
requires some effort to be solved. The first step consists in employing
integration by parts, which can be used once you notice that 

\[
\frac{\partial}{\partial\theta}\frac{1}{\left(\varepsilon+\cos\theta\right)}=\frac{\sin\theta}{\left(\varepsilon+\cos\theta\right)^{2}}.
\]
Therefore,

\begin{equation}
\int_{0}^{N\pi}\mathrm{d}\theta\;\frac{\sin^{2}\theta}{\left(\varepsilon+\cos\theta\right)^{2}}=\int_{0}^{N\pi}\mathrm{d}\theta\;\sin\theta\frac{\partial}{\partial\theta}\frac{1}{\left(\theta+\cos\theta\right)}=-\int_{0}^{N\pi}\mathrm{d}\theta\;\frac{\cos\theta}{\varepsilon+\cos\theta}.\label{eq:Parte1}
\end{equation}

The remaining integral can be simplified through the simple trick
of adding $\varepsilon-\left(-\varepsilon\right)$ to the numerator:

\begin{equation}
-\int_{0}^{N\pi}\mathrm{d}\theta\;\frac{\cos\theta}{\varepsilon+\cos\theta}=\int_{0}^{N\pi}\mathrm{d}\theta\;\frac{\varepsilon-\left(\varepsilon+\cos\theta\right)}{\varepsilon+\cos\theta}=\int_{0}^{N\pi}\mathrm{d}\theta\:\frac{\varepsilon}{\varepsilon+\cos\theta}-N\pi.\label{eq:Parte2}
\end{equation}
The cosine function is periodic, so the integrand in the extant integral
is the same in every period $\left[n\pi,\left(n+1\right)\pi\right]$,
for every $n$ from $0$ to $N-1$. Therefore,

\begin{equation}
\int_{0}^{N\pi}\mathrm{d}\theta\:\frac{\varepsilon}{\varepsilon+\cos\theta}=\sum_{n=0}^{N-1}\int_{n}^{\left(n+1\right)\pi}\mathrm{d}\theta\:\frac{\varepsilon}{\varepsilon+\cos\theta}=N\int_{0}^{\pi}\mathrm{d}\theta\:\frac{\varepsilon}{\varepsilon+\cos\theta}.\label{eq:Parte3}
\end{equation}

Now, we change variables to $w=\tan\left(\theta/2\right)$. This results
in the following differential:

\[
\frac{1}{2}\mathrm{d}\theta=\frac{\mathrm{dw}}{1+w^{2}},
\]
and allows us to express the cosine as follows:

\[
\cos\theta=\cos^{2}\left(\frac{\theta}{2}\right)-\sin^{2}\left(\frac{\theta}{2}\right)=\frac{1-\tan^{2}\left(\theta/2\right)}{1+\tan^{2}\left(\theta/2\right)}=\frac{1-w^{2}}{1+w^{2}}.
\]
Therefore, the integral in Eq. (\ref{eq:Parte3}) becomes:

\[
\int_{0}^{\pi}\mathrm{d}\theta\:\frac{\varepsilon}{\varepsilon+\cos\theta}=2\int_{0}^{\infty}\mathrm{d}w\;\frac{\varepsilon}{\varepsilon\left(1+w^{2}\right)+\left(1-w^{2}\right)}=\frac{2\varepsilon}{\varepsilon-1}\int_{0}^{\infty}\mathrm{d}w\;\frac{1}{\left(w+w_{0}\right)\left(w-w_{0}\right)},
\]
where $w_{0}\equiv i\sqrt{\left(\varepsilon+1\right)/\left(\varepsilon-1\right)}$.
Note that the number in the square root will always be greater than
zero, because $\varepsilon>1$ in both occasions treated in this article
(in the second integral, it represents the inverse of the eccentricity,
which is always a number between zero and one for an ellipse).

Using partial fractions,

\[
\int_{0}^{\infty}\mathrm{d}w\;\frac{1}{\left(w-w_{0}\right)\left(w+w_{0}\right)}=\frac{1}{2w_{0}}\left(\int_{0}^{\infty}\mathrm{d}w\;\frac{1}{w-w_{0}}-\int_{0}^{\infty}\mathrm{d}w\;\frac{1}{w+w_{0}}\right)=\frac{1}{2w_{0}}\int_{-\infty}^{\infty}\mathrm{d}w\:\frac{1}{w-w_{0}}.
\]
Added to the integral from $0$ to $\infty$, we have an integral
over the whole real line:

\begin{equation}
\int_{0}^{\pi}\mathrm{d}\theta\:\frac{\varepsilon}{\varepsilon+\cos\theta}=\frac{\varepsilon}{\varepsilon-1}\frac{1}{w_{0}}\int_{-\infty}^{\infty}\mathrm{d}w\;\frac{1}{w-w_{0}}.\label{eq:Parte4}
\end{equation}

\begin{figure}
\includegraphics{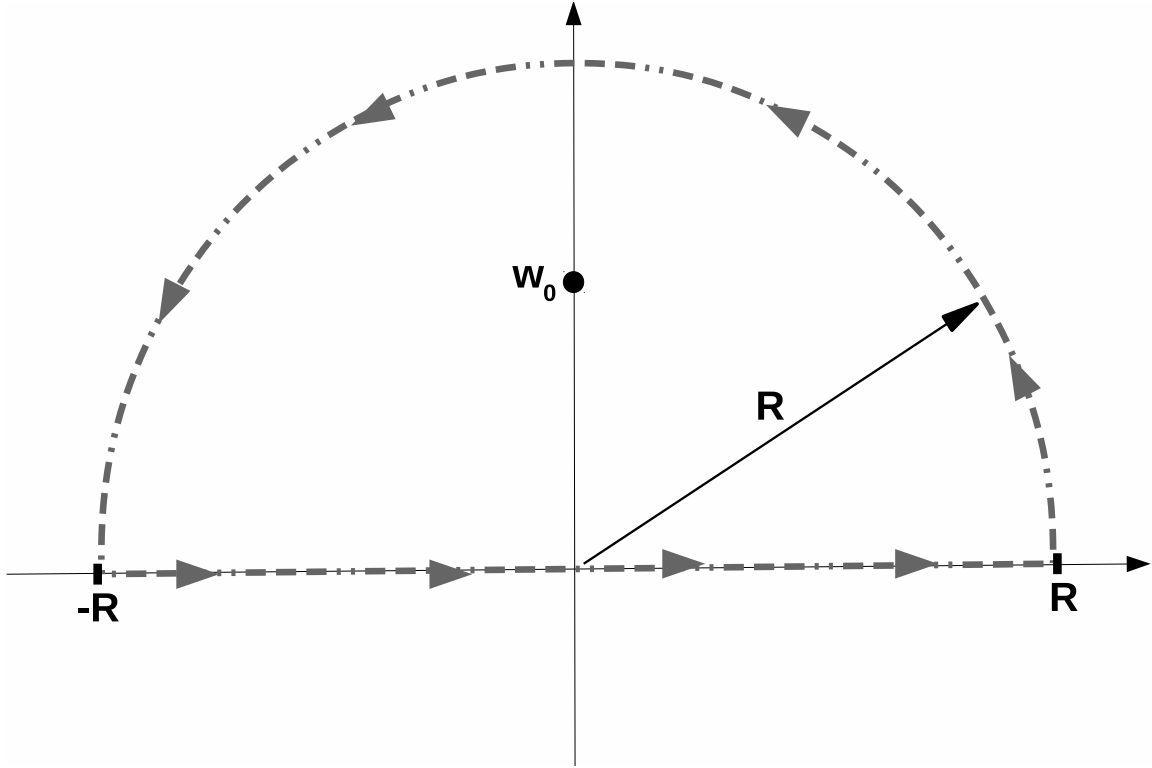}

\caption{Path of the complex-plane integral from Eq. (\ref{eq:PathIntegral}).
The upper arch is the integral of $\left(w-w_{0}\right)^{-1}$ when
$w=Re^{i\theta}$, with $\theta$ varying from $0$ to $\pi$. The
integral over the real line corresponds to integrating $\left(w-w_{0}\right)^{-1}$
from $-R$ to $R$, which in the limit when $R\to\infty$ is what
we want to calculate. Together, these two add up to the circuit integral
$\oint\mathrm{d}w\:\left(w-w_{0}\right)^{-1}$, which equals $2\pi i$
times the residues inside the circuit, which for the single pole $w=w_{0}$,
is $1$.}
\end{figure}

What is the value of the integral of $w$ through the real line? We
can use Cauchy's residue theorem to perform this integration through
a semi-circle of infinite radius that encompasses $w_{0}$ (see Fig.
3):

\begin{equation}
\int_{-\infty}^{\infty}\mathrm{d}w\;\frac{1}{w-w_{0}}=2i\pi-\lim_{R\to\infty}\int_{0}^{\pi}\mathrm{d}\theta\;\frac{iRe^{i\theta}}{Re^{i\theta}-w_{0}}=2i\pi-i\int_{0}^{\pi}\mathrm{d}\theta=i\pi.\label{eq:PathIntegral}
\end{equation}
If the student is not acquainted with residues, they can understand
this result by integrating from $-R$ to $R$ and later taking the
limit $R\to\infty$. The integral is obviously a logarithm:

\[
\int_{-\infty}^{\infty}\mathrm{d}w\;\frac{1}{w-w_{0}}=\lim_{R\to\infty}\int_{-R}^{R}\mathrm{d}w\;\frac{1}{w-w_{0}}=\lim_{R\to\infty}\ln\left(\frac{R-w_{0}}{-R-w_{0}}\right)=\ln\left(-1\right).
\]
As $e^{i\pi}=-1$, it is possible to see why the result of the integral
is $i\pi$.

Replacing this result in Eq. (\ref{eq:Parte4}), as well as the definition
of $w_{0}$, we find

\[
\int_{0}^{\pi}\mathrm{d}\theta\:\frac{\varepsilon}{\varepsilon+\cos\theta}=\frac{\varepsilon}{\varepsilon-1}\frac{i\pi}{w_{0}}=\frac{\varepsilon\pi}{\sqrt{\varepsilon^{2}-1}}.
\]

Replacing this in Eq. (\ref{eq:Parte3}) and then in Eq. (\ref{eq:Parte2})
and in Eq. (\ref{eq:Parte1}), we find the general result

\[
\int_{0}^{N\pi}\mathrm{d}\theta\;\frac{\sin^{2}\theta}{\left(\varepsilon+\cos\theta\right)^{2}}=N\frac{\varepsilon\pi}{\sqrt{\varepsilon^{2}-1}}-N\pi.
\]
Then, integral (\ref{eq:IntSec3}), from Sec. 3, becomes, with $N=1$,

\[
\int_{-\pi/2}^{\pi/2}\mathrm{d}\vartheta\:\frac{\cos^{2}\vartheta}{\left(\varepsilon+\sin\vartheta\right)^{2}}=\pi\left\{ \frac{\varepsilon}{\sqrt{\varepsilon^{2}-1}}-1\right\} ,
\]
while for the integral (\ref{eq:IntSec4}), from Sec. 4, we have $N=2$,

\[
\int_{0}^{2\pi}\mathrm{d}\theta\;\frac{\sin^{2}\theta}{\left(\varepsilon^{-1}+\cos\theta\right)^{2}}=2\pi\left\{ \frac{1}{\sqrt{1-\varepsilon^{2}}}-1\right\} .
\]


\begin{thebibliography}{10}
\bibitem[42]{footnoteH}According to Ref. {[}20{]}, Planck chose the
letter $h$ exactly because this was usually used as the variable
whose limit was taken to zero $\lim_{h\to0}$ in calculus textbooks.
\bibitem[47]{FootnoteSommerfeld}These articles by Sommerfeld in Ref.
{[}10-11{]} have no accessible English translations, even though his
ideas are found in his book mentioned in Ref. {[}6{]} and in the translations
of his previous communications to the Bavarian Academy of Sciences
in Ref. {[}32{]} and Ref. {[}33{]}.
\bibitem[53]{FootnoteMass}If one prefers not to employ this approximation,
use the reduced mass of the system $\mu$ instead of the mass of the
electron $m$.
\bibitem[54]{FootnoteGaussian}Here we used Gaussian units, so no
cumbersome constants have to be carried in the calculations. If the
student is more used to SI units, they may replace $Z$ with $Z/\left(4\pi\epsilon_{0}\right)$.
\bibitem[64]{FootnoteSchwarzchild}Refs. {[}62-63{]} present not just
a little of the history, but also the complete original calculations
of the Stark effect and comparison with modern methods.
\end{thebibliography}
\end{document}